\documentclass{JHEP3}
\usepackage[xdvi]{graphicx}
\usepackage{amsmath}
\usepackage{enumerate}

\title{Spacetime Embedding Diagrams for Spherically Symmetric Black Holes}
\author{ John T. Giblin, Jr$^*$, Donald Marolf$^\dagger$, and Robert Garvey$^*$\\
$^*$   College of the Holy Cross, Worcester, Massachusetts 01610 \\
$^\dagger$ Physics Department, Syracuse University, Syracuse, New York
13244 USA}

\abstract{
We show that it is possible to embed the 1+1 dimensional reduction of certain spherically symmetric black hole spacetimes into 2+1 Minkowski space. The spacetimes of interest (Schwarzschild de-Sitter, Schwarzschild anti de-Sitter, and Reissner-Nordstrom near the outer horizon)  represent a class of metrics whose geometries allow for such embeddings.  The embedding diagrams have a dynamic character which allows one to represent the motion of test particles.  We also analyze various features of the embedding construction, deriving the general conditions under which our procedure provides a smooth embedding. These conditions also yield an embedding constant related to the surface gravity of the relevant horizon.
}

\begin{document}


\section{Introduction}

\begin{figure}[h] 
\centerline{\includegraphics{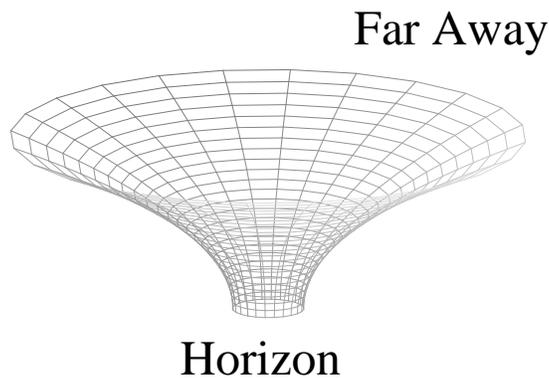}}
\caption{The spatial exterior $r$-$\varphi$ plane of a Schwarzschild Black Hole for $t=constant$.
\label{spaceEmb}} 
\end{figure}

Embedding diagrams are excellent tools for presenting complicated curvature information in a simple visual format.  In general relativity, they can be used to aid the development of intuition needed to deal with curved spacetimes.  They can thus be extremely useful tools for educators.  However, such diagrams (as in figure \ref{spaceEmb}) are often used to convey only information about the spatial curvature of constant-time slices, as opposed the spacetime curvature information that controls, for example, the relative acceleration of nearby geodesics.

\begin{figure}[h]
\centerline{\resizebox{!}{5cm}{\includegraphics{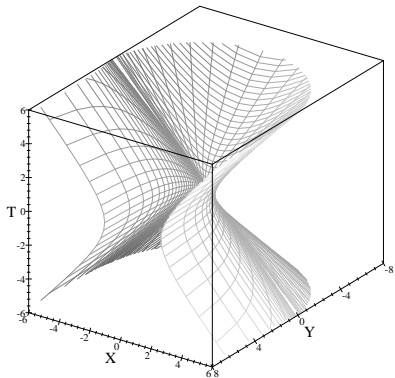}}}
\caption{Embedding Diagram of Kruskal Black hole with $r$ and $t$ coordinates. \label{SS1}}
\end{figure}

\smallskip
In the above diagram the equatorial plane (the radial and one angular dimension) of a Schwarzschild black hole is embedded in three dimensional Euclidean space.  This type of embedding conveys a certain amount of information about the spatial curvature of the {\it exterior} of the black hole, but it includes neither information about spacetime curvature nor in fact about any curvature in the interior.  In contrast, diagrams such as figure \ref{SS1} below were presented in \cite{ma} and display spacetime curvature by showing the $r,t$ plane (i.e., the slice $\theta,\varphi= constant$) of the Kruskal extension embedded in 2+1 Minkowski space.   This diagram was explained in \cite{ma} and an overview will be recalled below.
Here we use the technique of \cite{ma} to create such embedding diagrams for a variety of black hole spacetimes.  For earlier work involving the embedding of the associated 3+1 spacetimes into higher dimensional flat spaces see \cite{Kasner,Fronsdal, Deser,Pad1,Pad,Davidson}

Note that figure \ref{spaceEmb} relies on a 2-dimensional reduction where both remaining dimensions are spacelike.  In order to depict space{\it time} curvature, it is natural to embed a $(1+1)$ dimensional reduction into Minkowski space.  In particular, this was done in figure \ref{SS1}.  Embeddings of totally geodesic surfaces containing a timelike direction are dynamic in the sense that they are capable of displaying timelike geodesics and other particle trajectories \cite{ma}.

Such diagrams are appropriate for an audience with knowledge of special relativity and familiarity with both Minkowski space and $(1+1)$ spacetime diagrams.  The diagrams may be used to educate students at this level about general relativity and curved spacetime, or they can provide an additional perspective for physicists already familiar with the subject.  We are particularly interested in properties of black hole horizons.

The diagrams described here include regions on both sides of a horizon.  We describe the general construction and derive conditions for its success in section 2.  We then construct diagrams for the de-Sitter, Schwarzschild-de Sitter,  Schwarzschild-anti de Sitter, and Reissner-Nordstrom spacetimes.  These cases were chosen in order to explore the embedding process for spherically symmetric spacetimes with multiple horizons.  We will see that in general our construction is unable to smoothly embed a region of the spacetime containing both horizons.  Instead, one must embed the spacetime in patches containing only one horizon.  For the Reissner-Nordstrom case, a smooth embedding is only obtained for the outer horizon, and even then only when $Q < \sqrt{8/9} M$, as defined in section \ref{rnst}. 

\section{The Embedding Process}
\label{process}

The embedding process is a generalization of that described in Appendix A of \cite{ma} for the Kruskal black hole.  For this reason, we will refer to the Kruskal case often as a familiar point of reference.  For spacetimes with more than one horizon, we will focus on each horizon separately.

We have chosen a class of 3+1 spacetimes that are both spherically symmetric and static, at least outside of some horizon.  The spherically symmetric condition makes it possible to set $\theta=const$ and $\varphi=const$ to obtain a 1+1 dimensional reduction which retains significant geometric information.  In particular, such a slice is a totally geodesic surface, meaning that geodesics in the induced metric on the slice coincide with geodesics in the full spacetime.  

Let us consider metrics of the form:

\begin{equation}
\label{symmet}
ds^2=-\phi dt^2+\phi^{-1} dr^2+r^2d\Omega^2.
\end{equation}
\noindent
where $d\Omega^2 = d\theta^2+sin^2\theta d\varphi^2$.  For our purposes, the metric can be reduced to
\begin{equation}
\label{dimred}
ds^2=-\phi dt^2+\phi^{-1} dr^2.
\end{equation}
\noindent
Since we have restricted ourselves to static metrics, $\phi$ is a function of $r$ only.  We henceforth use units such that both the speed of light, $c$, and Newton's constant, $G$, are equal to one.  Although (\ref{symmet}) is not the most general form of a spherically symmetric metric, it represents a class of metrics for which the energy density is equal to the radial pressure, $\rho = p$ ($T^1_1 = T^0_0$)---a class that includes many interesting examples.

As long as $\phi > 0$, changes in $t$ are timelike and the spacetime is indeed static.  If, however, at any point $\phi=0$ we encounter a coordinate singularity similar to that seen in a static coordinate system at the horizon of a black hole.  Such locations are clearly Killing horizons, as $|\partial_t|$ vanishes.  One may verify that this is indeed a smooth horizon so long as $\phi$ is a smooth function of $r$ at the coordinate singularity.  The horizon is degenerate if and only if $\partial_r \phi =0$.  We will see that our method yields an embedding when $\partial_r \phi \neq 0$ and $\partial_r^2 \phi < 0$ at the horizon.

Recall that we wish to embed our reductions into $(2+1)$ Minkowski space with metric:

\begin{equation}
\label{2+1Mink}
ds^2=-dT^2 + dX^2 +dY^2.
\end{equation} 
We break up Minkowski space into four sections, defined by two intersecting planes $X=\pm T$ and numbered I, II, III, and IV as in Fig. 2 (with the coordinate Y coming out of the page).

\smallskip

\begin{figure}[h]
\centerline{\includegraphics{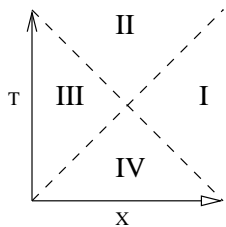}}
\caption{Division of $(2+1)$ Minkowski Space \label{fig:Mink}}
\end{figure}
\smallskip
\noindent

We define a system of hyperbolic cylindrical coordinates $(\rho, \psi, Y)$ in each section.  These coordinates are so named because $Y$ will remain a linear coordinate, as in cylindrical coordinates.  For region I these are:

\begin{eqnarray}
\nonumber \rho &=& \sqrt{X^2-T^2} \\
\nonumber \psi &=& \tanh^{-1} \left({T \over X}\right).
\end{eqnarray}

\noindent
In region II, they become:

\begin{eqnarray}
\nonumber \rho &=& \sqrt{T^2-X^2} \\
\nonumber \psi &=& \tanh^{-1} \left({X \over T}\right).
\end{eqnarray}

Coordinates in regions III and IV can be obtained by reflection of regions I and II.
In region I, the metric (\ref{2+1Mink}) becomes

\begin{equation}
\label{hycl}
ds^2=-\rho^2 d \psi^2 + d \rho^2 + dY^2.
\end{equation}
This is one illustration of the well-known fact that the planes $X = \pm T$ are horizons
in Minkowski space. 

\begin{figure}[h]
\centerline{\includegraphics{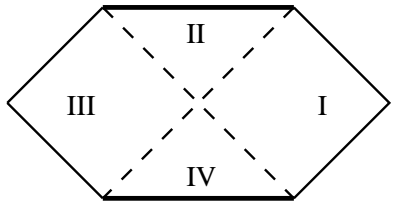}}
\caption{The Penrose Diagram for the Kruskal spacetime\label{fig:S}}
\end{figure}

Suppose now that one compares this with a 1+1 spacetime of the form (\ref{dimred}) near a zero of $\phi$.  With foresight let us suppose that $\partial_r \phi ={ 2 \kappa}$ at this zero, so that in the leading approximation we have $\phi = {2 \kappa r}$.  Then if we introduce $\rho = {1 \over \kappa } \sqrt{2 \kappa r}$ and $\psi = {t \kappa}$, the metric (\ref{dimred}) takes the form $ds^2=-\rho^2 d \psi^2 + d\rho^2$, matching the first two terms of (\ref{hycl}).  This is just the usual result that any smooth non-degenerate horizon is locally like the above horizon in 1+1 Minkowski space; i.e. in any slice $Y=constant$.  This similarity is apparent in the familiar Penrose diagram (figure \ref{fig:S}) for the Kruskal spacetime.  We refer to the quadrants surrounding any singularity of this nature as I,II,III,IV in analogy with figure \ref{fig:Mink}, though for the remainder of this section we will speak only of region I unless otherwise noted.

Our intention is to go beyond this leading approximation and achieve an exact equality over some finite region of the spacetime by comparing (\ref{dimred}) with some more interesting hypersurface $Y = Y(\rho, \psi)$.  

It is clear that any metric of the form (\ref{dimred}) has a Killing field corresponding to a $t$ translation; in other words that a transformation of the form  $(t,r) \rightarrow (t + \delta t,r)$ is length preserving.  In (\ref{dimred}), the timelike distance between two events along a curve of constant $r$ separated by $\delta t$ is $\delta t \sqrt{\phi}.$  However in Minkowski space (\ref{hycl}) a boost $(\psi, \rho, Y)  \rightarrow (\psi + \delta \psi, \rho, Y)$ translates points by $\rho \delta \phi$.  To match these Killing fields we must again set 
\begin{equation}
\psi = {t \kappa}
\end{equation}
\noindent
for some constant $\kappa$.  Since the above transformations must move points through the same proper time, we must also have 
\begin{equation}
\label{rhophi}
\rho = {1 \over \kappa}  \sqrt{\phi}.
\end{equation}

The particular surface $Y = Y(\rho,\psi)$ is then determined by setting the full distance elements equal.  After our identifications above, this is equivalent to requiring agreement on a slice of constant $t = \psi/\kappa$:  
\begin{equation}
 d \rho^2 +dY^2 =\phi^{-1} dr^2.
\end{equation}
Using equation (\ref{rhophi}), one may solve for $dY$:
\begin{equation}
\label{dYsol}
dY = \pm  \sqrt{1-{1 \over 4 \kappa^2 }\left({d \phi \over dr}\right)^2 \over \phi} dr.
\end{equation}
For convenience, we take $Y=0$ at the horizon $r = r_h$.  In region I we will choose the positive
sign to yield
\begin {equation}
\label{Ysol}
Y=\int_{r_h}^r{ \sqrt{1-{1 \over 4 \kappa^2 }\left({d \phi \over dr}\right)^2 \over \phi} dr}.
\end{equation}

Proceeding similarly in region II one again arrives at exactly\footnote{One might have expected that additional minus signs would arise in region II, but this turns out not to be the case in the final expression for $Y$.} (\ref{dYsol}) and (\ref{Ysol}).  Now, $r$ is a smooth function on the spacetime so, if the if the embedding is to be smooth (and in particular if $dY/dr$ is to be real) we must require that $1-{1 \over 4 \kappa^2 }\left({d \phi \over dr}\right)^2$ change signs when $\phi$ does; i.e., at the horizon.  Thus we find
\begin{equation}
\label{ec}
\kappa =  \frac{1}{2}  \left.{d\phi \over dr}\right|_{r=r_{h}} ,
\end{equation}
which is exactly what was used in the leading approximation above.
One could also take $\kappa$ to have the opposite sign, but it is useful for us to fix a convention using (\ref{ec}).

We should note that our notation does not match that of \cite{ma}, where our constant $\kappa$ was called $1/\kappa$.  To interpret this constant, recall \cite{wa} that the surface gravity of a Killing horizon is given by the expression:
\begin{displaymath}
\kappa^2=-{1 \over 2} (\nabla^a\xi^b)(\nabla_a\xi_b),
\end{displaymath}
\noindent
where $\xi^a$ is the stationary Killing field.  Evaluating this for the metric (\ref{dimred}) for $\xi$ a time translation one finds that the surface gravity is 
\begin{equation}
\label{sg}
\kappa= \pm  \frac{1}{2}  \left.{d\phi \over dr}\right|_{r=r_h}.
\end{equation}

By convention, the surface gravity is usually taken positive, but in any case we see that it agrees with
our embedding constant $\kappa$ up to the above sign.
Since the embedding constant $\kappa$ is unique, we will be unable to embed any section of a spacetime having several horizons with distinct values of $|\partial_r \phi|$.   However, we will see below that such spacetimes can typically be embedded in patches, with each patch containing only one horizon.

Finally, in order to yield an embedding, the quantity
\begin{displaymath}
{1-{1 \over 4 \kappa^2 }\left({d \phi \over dr}\right)^2 \over \phi}
\end{displaymath}
under the square root in the integrand of (\ref{Ysol}) must be positive in a neighborhood of the horizon.  
Near the horizon we may use the approximations

\begin{eqnarray}
\nonumber {d \phi \over dr} &=& \left({d \phi \over dr}\right)_{r=r_h}
+\left({d^2 \phi \over dr^2}\right)_{r=r_h}(r-r_h)+O\left( (r-r_h)^2 \right )\\
\nonumber &=& 2 \kappa+\left({d^2 \phi \over dr^2}\right)_{r=r_h}(r-r_h)+O\left( (r-r_h)^2 \right ), \\
\end{eqnarray}

\noindent
and 

\begin{displaymath}
\phi = 0 + 2 \kappa (r-r_h) + O\left( (r-r_h)^2 \right ).
\end{displaymath}

Substitution of these expressions into the integrand of (\ref{Ysol}) yields:

\begin{displaymath}
\sqrt{-  \frac{1}{2\kappa^2}  \left({d^2 \phi \over dr^2}\right)\Bigg|_{r=r_h} + O(r-r_h)},
\end{displaymath}

\noindent
so that we produce an embedding when

\begin{equation}
\label{nsd}
{d^2 \phi \over dr^2} \big|_{r=r_h} <0.
\end{equation}

In the complimentary case $\partial_r^2 \phi |_{r=r_h} > 0$, the argument of the square root is negative on both sides of the horizon for $\kappa = \frac{1}{2} \partial_r \phi$.  By choosing another value for $\kappa$, one may embed a neighborhood of any point with $\phi \neq 0$, but our method will not produce a smooth embedding of the horizon itself.  In either case, 
embeddings can fail at points when $\partial_r \phi=\pm 2 \kappa$ as $1-{1 \over 4 \kappa^2 }\left({d \phi \over dr}\right)^2$ may change sign. 

For a Kruskal black hole one finds $\kappa={1 \over 4M}$ and the complete embedding (in region I) is:
\begin{eqnarray}
\nonumber \rho&=&{1 \over 4M} \sqrt{1-{2M \over r}}\\
\nonumber \phi&=& { 4M t}\\
\nonumber Y&=&\int_{2M}^r \sqrt{\left({1+{2M \over r}+{4M^2 \over r^2}+{8M^3 \over r^3}}\right)}
\end{eqnarray}
as described in \cite{ma}.  
The embedding equations for the other three regions differ only by occasional negative signs.  The resulting embedding diagram was already displayed  in figure \ref{SS1} and is explained in detail in \cite{ma}.

\section{Spacetimes with a cosmological constant}

In seeking interesting examples beyond the Kruskal black hole, it is natural to investigate spacetimes with a cosmological constant.  In this section, we construct the $rt$ embedding diagrams for the Schwarzschild-de Sitter and Schwarzschild-Anti de Sitter spacetimes.  First, however, it is of interest to examine how our techniques apply to de Sitter (dS) and anti-de Sitter spaces (AdS) themselves.

\begin{figure}[h]
\label{ds}
\centerline{%
\resizebox{!}{6cm}{\includegraphics{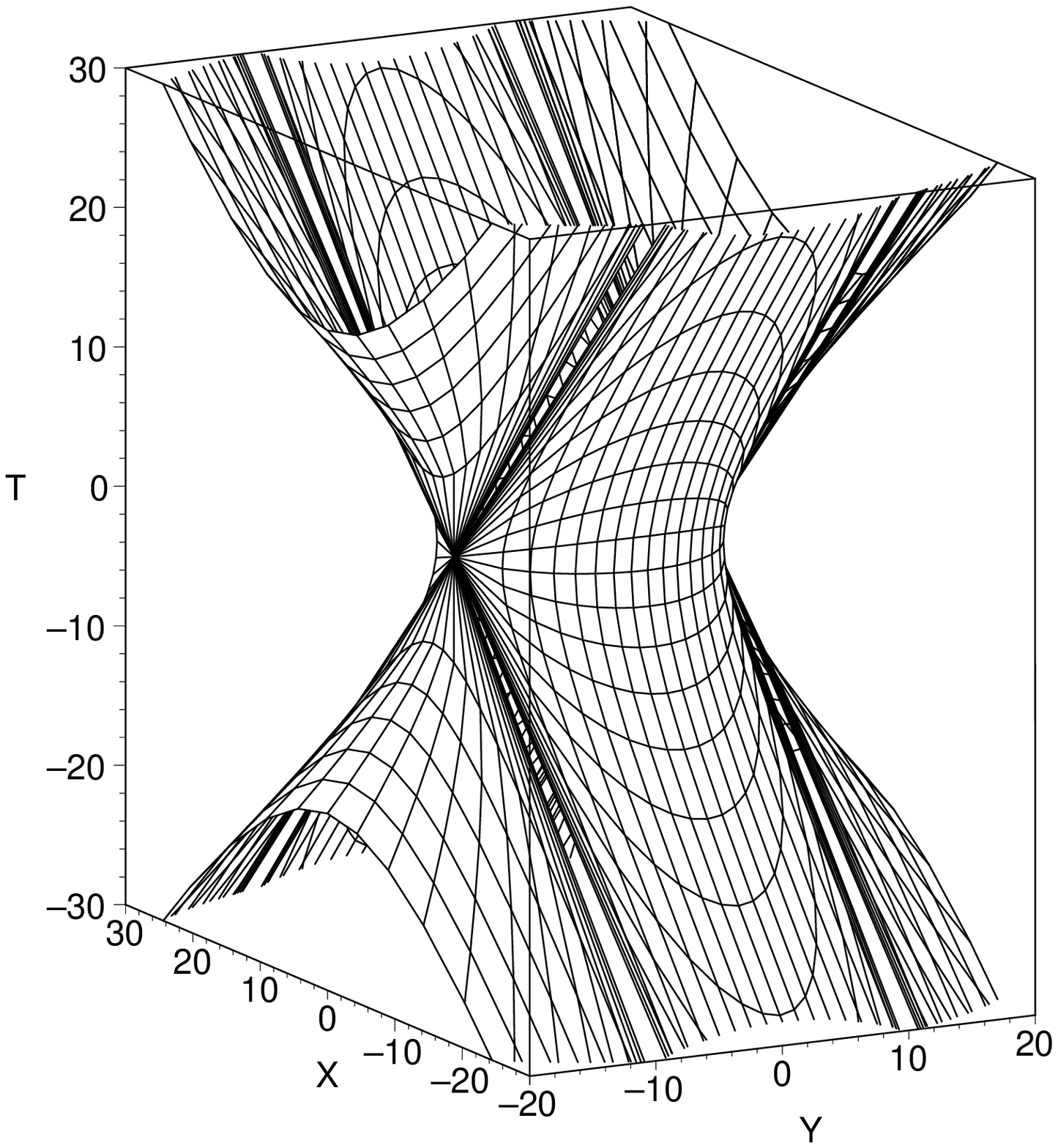}}
\resizebox{!}{6cm}{\includegraphics{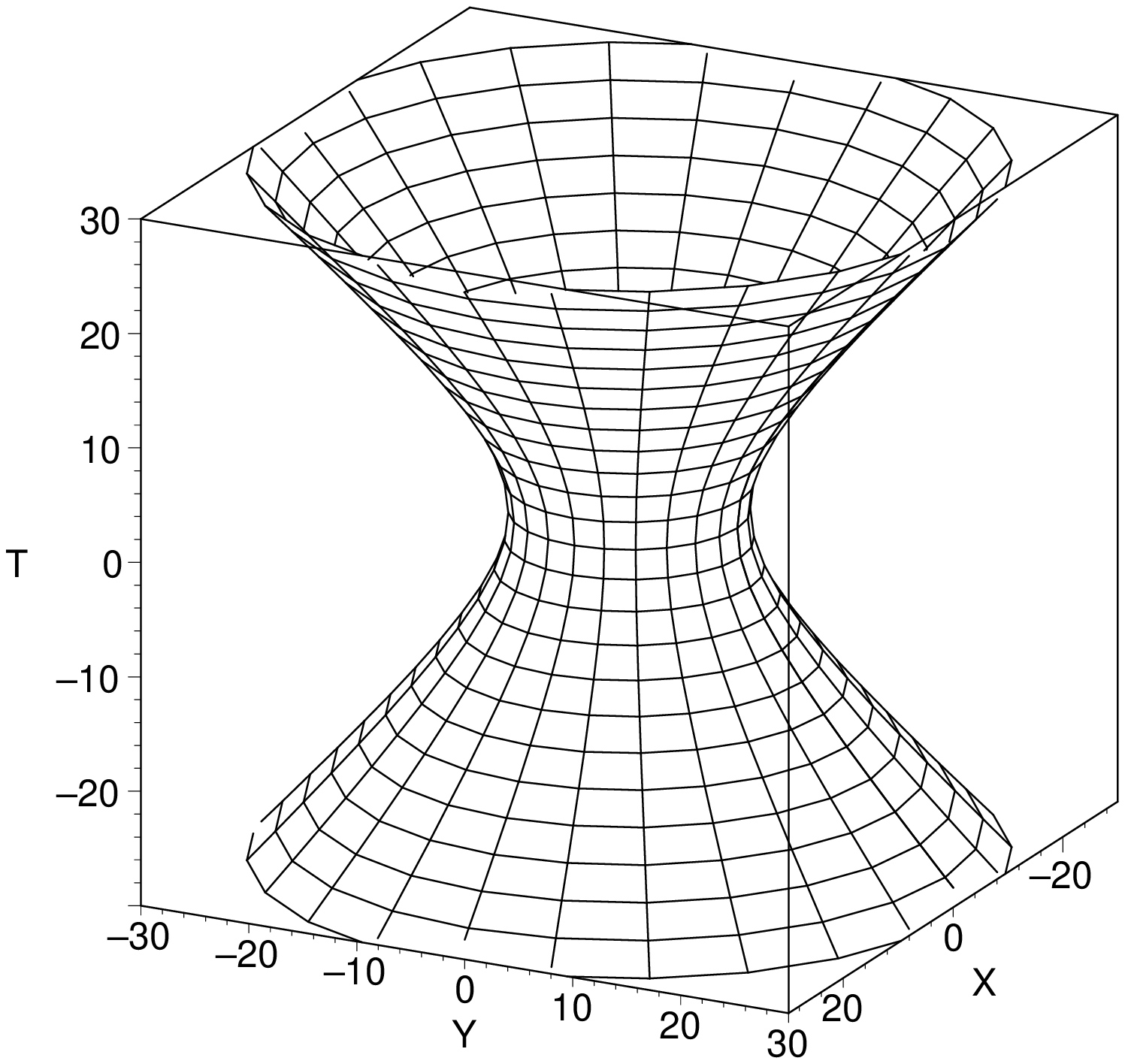}}}
\caption{De-Sitter space (with $a=10$) in terms of six patches with time or space translational symmetry (left) and in the standard presentation (right).  The above diagrams represent double covers of the $rt$ plane as defined in the text, with the same values of $r$ being shown at both positive and negative $Y$.
\label{2ds}
}
\end{figure}

Let us begin with de Sitter space.  In static coordinates, the metric takes the form \cite{ri}

\begin{displaymath}
ds^2=-\left(1-{r^2 \over a^2} \right)dt^2+\left(1-{r^2 \over a^2}\right)^{-1}dr^2+r^2(d\theta^2+sin^2\theta d\varphi^2),
\end{displaymath}
where $a^{-2} = \Lambda/3$ and $\Lambda$ is the usual cosmological constant.  Thus, $\phi = 1 - \frac{r^2}{a^2}$.  Since $\partial_r \phi \neq 0$ and $\partial^2_r \phi < 0$ at $r=a$, we may apply the method of section \ref{process}.  The surface gravity is $\kappa = -1/a$, and the resulting diagram is shown on the left in figure \ref{ds}.  In our method, the full diagram is formed by patching together 6 distinct regions.  However, the end result is clear from the fact that the 1+1 dimensional reduction of dS$_{n+1}$ is just 1+1 de Sitter space\footnote{This result is in turn clear from the representation of dS$_{n+1}$ as a hyperboloid in (n+1)+1 Minkowski space.}.  The usual embedding diagram of 1+1 de Sitter is shown on the right in figure \ref{ds}.

\begin{figure}[h]
\label{ads}
\centerline{\resizebox{!}{4cm}{\includegraphics{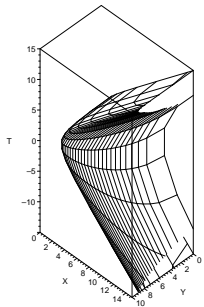}}}
\caption{Anti de-Sitter Spacetime with $a=10$ and $\kappa =1$}
\end{figure}

On the other hand, one might consider anti de-Sitter space (characterized by a negative cosmological constant) in the static form:

\begin{equation}
\label{adsmetric}
ds^2=-\left({1+{r^2 \over a^2}}\right) dt^2 + {1 \over \left({1+{r^2
 \over a^2}}\right)} dr^2 + r^2 d\Omega^2.
\end{equation}
The translation $\partial_t$ has no Killing horizon, so there is no preferred value of the embedding constant $\kappa$.
However, comparing (\ref{dYsol}) and (\ref{adsmetric}) we see that we can embed the surface only for $\frac{r^2}{a^4} < 4 \kappa^2$, at which point $\frac{dY}{dr}$ vanishes.  The result is shown in figure \ref{ads}.

\subsection{Schwarzschild de-Sitter Spacetime}

Let us now add a black hole to de Sitter space.  Using the conventions above, the metric for the radial plane of the resulting Schwarzschild-de Sitter spacetime is given by

\begin{equation}
\label{sdsmetric}
ds^2= -\left({1-{2M \over r}-{r^2 \over a^2}}\right)dt^2+\left({1-{2M \over r}-{r^2 \over a^2}}\right)^{-1}dr^2,
\end{equation}

\noindent
where $\phi = 1-{2M \over r}-{r^2 \over a^2}$.  This $\phi$ has an even number of positive real zeros representing horizons and an odd number of (unphysical) negative real zeros. The case with no horizons is not of interest, so we focus on the case with two horizons at $r_{BH}$ and $r_{cosmo}$ with $r_{BH} < r_{cosmo}$.  These are usually interpreted as the black hole and cosmological horizons respectively.  Since the second derivative $\partial_r^2 \phi$ is negative for all $r>0$, a neighborhood of either horizon may be embedded by our technique.

We rely upon a numerical solution to produce the final diagrams, and we choose $a=10$ and $M=1$ as illustrative values.  It follows that $r_{BH} = 2.091488484 $ and $ r_{cosmo} = 8.788850662$.  Since there are two horizons we will need two patches to completely represent the spacetime.  Using (\ref{ec}), we can easily find both values of $\kappa$:

\begin{eqnarray}
\kappa_{BH} &=& .2076918626 \\
\kappa_{cosmo} &=& -.07494249985
\end{eqnarray}
We refer to these patches as the ``black hole patch'' and the ``cosmological patch'' according to which horizon they smoothly embed.  Note that $|\frac{\kappa_{BH}}{\kappa_{cosmo}}|>1$, a condition true for all non-extremal choices of $a$ and $M$.  As a result the black hole patch can extend all the way to the cosmological horizon without encountering a point where $\partial_{r} \phi = 2\kappa_{BH}$. The cosmological patch, in contrast, will always encounter a point where $\partial_{r} \phi = 2 \kappa_{cosmo}$ and it will fail to extend to the black hole horizon.

\smallskip
\begin{figure}[h]
\centerline{%
\resizebox{!}{5cm}{\includegraphics{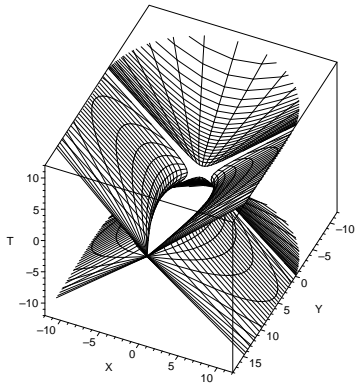}}}
\caption{The black hole patch of Schwarzschild de-Sitter spacetime. \label{sdsin}}
\end{figure}

The embedding is straightforward.  The black hole patch is shown in figure \ref{sdsin}.  The region around the black hole horizon resembles that of the Schwarzschild case, while the intermediate regions (I and III) pinch off to form crossed null planes as one approaches the cosmological horizon.   This demonstrates that $r=r_{cosmo}$ is a null surface, though we cannot demonstate
the smoothness of this surface.  The null planes can be seen analytically as, when $\partial_t$ is timelike, the tangent line to the intersection of our surface with any constant $Y$ plane has slope

\begin{displaymath}
{dT \over dX} = {{dT \over dt} \over {dX \over dt}} = \coth(\kappa t).
\end{displaymath} 
\noindent
As one approaches any point on the black hole horizon away from the bifurcation surface, one has $t\rightarrow \pm \infty$ so that the slope of our $Y=const$ slices approaches $\pm 1$.

\smallskip
\begin{figure}[h]
\centerline{%
\resizebox{!}{5cm}{\includegraphics{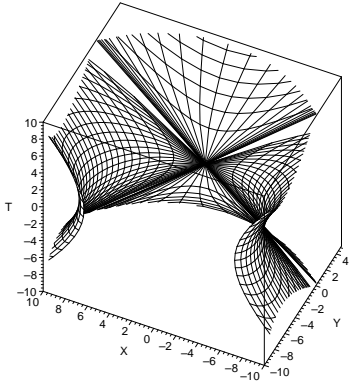}}
}
\caption{The cosmological patch of Schwarzschild de-Sitter spacetime shown from two perspectives in the embedding space.  This diagram terminates at $r=3.075820840$, where $\partial_r \phi = - 2 \kappa$.
 \label{sdsout}}
\end{figure}

 In the cosmological patch (figure \ref{sdsout}) of Schwarzschild de-Sitter, the diagram near the cosmological horizon has a structure similar to that of its de-Sitter space counterpart (figure \ref{ds}).  However, when one moves away from $y=0$ one reaches
a point where $\partial_r \phi =- 2 \kappa$ and the embedding terminates.

\begin{figure}[h]
\centerline{
\includegraphics{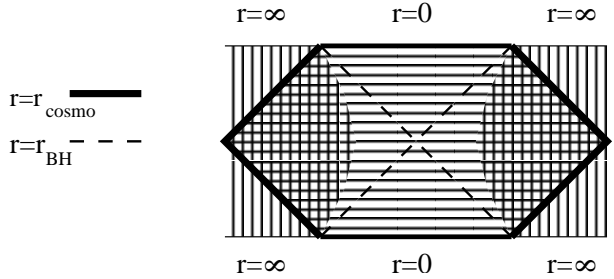}}
\caption{Penrose diagram for Schwarzschild de-Sitter spacetime. The cosmological patch is shaded horizontally, while the black hole patch is shaded vertically.  Note that there is a large overlap region which is contained in both patches.  It is natural to identify the left and right edges of this diagram.  Alternatively, the maximal analytic extension is achieved by taking the pattern to repeat infinitely to both left and right. 
\label{penrosesds}}
\end{figure}

We illustrate the two (overlapping) patches used by shading the corresponding parts of the Schwarzschild-de Sitter Penrose diagram in figure \ref{penrosesds}.

\smallskip
\smallskip

\subsection{Schwarzschild Anti de-Sitter Spacetime}

The addition of a black hole to Anti de-Sitter space creates a horizon of non-zero surface gravity.  The full (3+1) metric is:

\begin{equation}
ds^2=-\left({1-{2M/r}+{r^2 \over a^2}}\right) dt^2 + {1 \over
\left({1-{2M/r}+{r^2 \over a^2}}\right)} dr^2 + r^2 d\Omega^2.
\end{equation}

\begin{figure}[h]
\centerline{%
\resizebox{!}{5cm}{\includegraphics{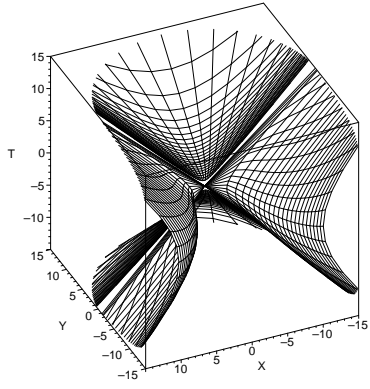}}}
\caption{Embedding Diagram for Schwarzschild Anti de-Sitter Spacetime
with $a=10$ and $M=1$.  The corresponding surface gravity is $\kappa=3.469567820$. This diagram terminates at $r=28.70063399$---the closest value of $r$ satisfying $\partial_r \phi= 2\kappa$ \label{sads}}
\end{figure}

\begin{figure}[h]
\centerline{\includegraphics{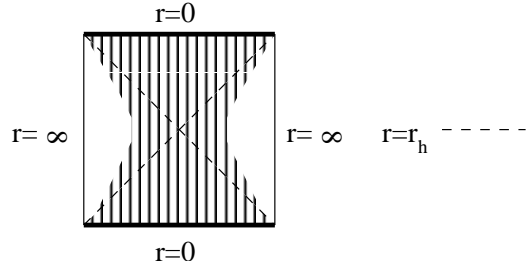}}
\caption{Penrose Diagram for Schwarzschild Anti de-Sitter Spacetime.  Vertical shading represents the portion of spacetime embedded in figure 10. \label{penrosesads}}
\end{figure}

Thus $\phi = 1-{2M \over r}+{r^2 \over a^2}$ which, for any $M$ and $a$, has a single positive real root and two complex roots.  One may check that $\partial_r^2 \phi < 0$ at this horizon.  Thus our construction will yield a smooth embedding, and the above patch will cover the spacetime out to the next value of $r$ for which $\partial_r \phi = \pm 2 \kappa$, which exists (for $r> r_h$)
for any value of $2M$ and $a$.
The result is shown in figure \ref{sads}.  The Penrose diagram for this spacetime is shown in figure \ref{penrosesads}.

\section{Reissner-Nordstrom Spacetime}
\label{rnst}
The Reissner-Nordstrom spacetime is a black hole with an electric (or magnetic) charge $Q$.  The addition of this charge changes the Schwarzschild solution to

\begin{displaymath}
ds^2 = -\left({1-{2M \over r}+{Q^2 \over r^2}}\right) dt^2+\left({1-{2M \over r}+{Q^2 \over r^2}}\right)^{-1} dr^2 + r^2\left({d\theta^2 + sin^2\theta d\varphi^2}\right),
\end{displaymath}

\noindent
where $\phi = \left({1-{2M \over r}+{Q^2 \over r^2}}\right)$.  Again, $\phi$ will have two roots.  These  occur at $r_{+}=M+\sqrt{M^2-Q^2}$ and $r_{-}=M-\sqrt{M^2-Q^2}$.    This time, however, $\partial_r^2 \phi$ is not of  definite sign.  It is negative for $r > {3Q^2 \over 2M}$, but positive for $r < {3Q^2 \over 2M}$. The value  $r = \frac{3Q^2}{2M}$ is always larger than $r_-$, and becomes larger than $r_+$ for $Q > \sqrt{8/9} M$.  Thus we can smoothly embed only the outer horizon\footnote{As usual, an appropriate choice of $\kappa$ will allow us to locally embed the surface at any
point for which $\phi \neq 0$, but the corresponding patch will never contain the inner horizon.}, and that only for black holes not too close to extremality.  

As one might expect in this situation, a smoothly embedded patch around the outer horizon will never be able even to approach the
inner horizon.  Thus will be the case whenever one considers two horizons separated by a region where $\partial_t$ is
timelike.  To see this, note that $\phi$ must have a minimum between the two horizons, where of course $\partial_r \phi =0$.
But at the outer horizon we have $\partial_r \phi = 2\kappa > 0$ and $\partial^2_r \phi > 0$.  Thus, moving inward from the outer
horizon $\partial_r \phi$ must increase at first and then later decrease to zero before one approaches the inner horizon.  
Somewhere along the way $\partial_r \phi$ takes on the value $2\kappa$ and the embedding terminates.  The same would be true
for a patch that smoothly embeds a region near the inner horizon (though, as stated above, none exists for Reissner-Nordstrom).

\begin{figure}[h]
\centerline{%
\resizebox{!}{5cm}{\includegraphics{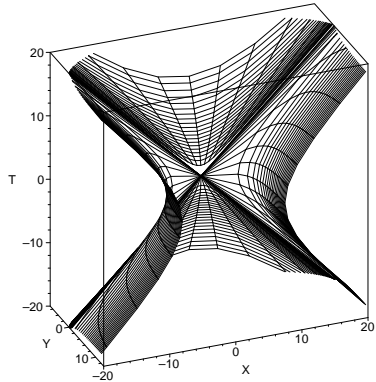}}}
\caption{A patch of Reissner-Nordstrom Spacetime ($M=2=2Q$). The embedding ends at $r=.508159$ where $ \partial_r \phi=  2\kappa$ \label{rnout}.}
\end{figure}

\begin{figure}[h]
\centerline{%
\includegraphics{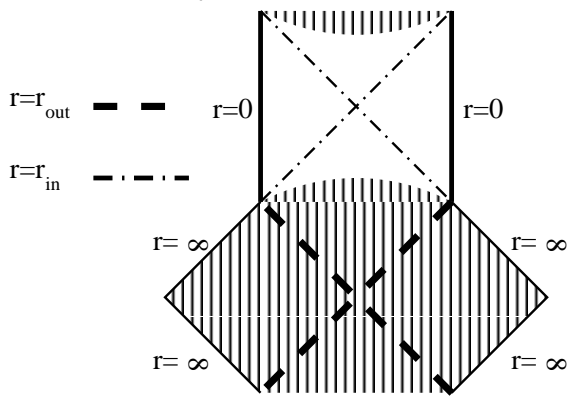}}
\caption{Penrose diagram for the extended Reissner-Nordstrom spacetime.  Vertical shading represents the patch surrounding $r=r_{+}$. \label{penrosern22}}
\end{figure}

In contrast, our patch does yiel;d a smooth embedding for all $r > r_+$.
An associated diagram (for $Q < \sqrt{8/9} M$)  is shown in figure \ref{rnout}.  Upon initial inspection, the outer patch of this embedding (figure \ref{rnout}) looks much like the Schwarzschild black hole.  The Penrose diagram (figure \ref{penrosern22}) for the Reissner-Nordstrom case also exhibits a periodic structure but this time in a timelike direction.  Again, we have shaded this diagram to indicate the region shown in figure \ref{rnout}.

\section{Discussion} 

In the above work we have analyzed the embedding process of \cite{ma} in detail.  This procedure embeds spherical reductions of spherically symmetric static spacetimes into 2+1 Minkowski space. The resulting embedding diagrams directly show the behavior of radial geodesics (both timelike and spacelike) and other dynamic worldlines in the original spacetime.  

We have constructed a number of examples which illustrate the issues involved in constructing embedding diagrams for spacetimes with more than one horizon.  The technique generically does not embed patches containing more than one horizon, though in some cases a patch containing one horizon can approach a second horizon.  When this occurs, the embedding diagram makes the null character of the static Killing field clear at the second horizon.  What is unclear from the diagram is that the spacetime is smooth at this second horizon.  We have also analyzed conditions under which an embedding terminates before reaching a second horizon.

Most importantly, we have analyzed the embedding process in detail and found that, for metrics of the form (\ref{symmet}), a patch containing a horizon (i.e., a zero of $\phi$) can be embedded by this technique when  $\partial_r \phi \neq 0$ and $\partial_r^2 \phi < 0.$  This result is easily generalized to the generic case.  An arbitrary static spherically symmetric metric can be written in the form

\begin{equation}
\label{arbmetric}
ds^2 = g_{tt} dt^2 + g_{rr} dr^2 + r^2 (d\theta^2 + \sin^2 \theta d\psi^2).
\end{equation}
Repeating the calculations of section \ref{process} for the metric (\ref{arbmetric}),
one finds that the embedding becomes
\begin{equation}
\psi = {t \kappa}, \ \ \ 
\rho = {1 \over \kappa}  \sqrt{-g_{tt}}, \ \ \ 
Y=\int_{r_h}^r\sqrt{g_{rr} + {1 \over 4 \kappa^2 g_{tt} }\left({d g_{tt} \over dr}\right)^2 }dr,
\end{equation}
with the conditions for success near a given horizon being that, again, $\kappa \neq 0$ must be the surface gravity and that
$\partial_r^2 g_{tt}$ must be positive.

The reader will note that we can considered the most commonly discussed black hole solutions with the exception of the
extremal cases.  Extreme black holes have vanishing surface gravity, which is incompatible with the explicit construction
given here.  This is due to our mapping the black hole horizon to a bifurcate horizon in 2+1 Minkowski space.
One may be able to obtain an embedding of the region near e.g. the future horizon of
some extreme black holes by starting with our procedure and taking a limit.  The results of section \ref{rnst}
indicate that this is unlikely  in the Reissner-Norstrom case, but the exploration of other extreme black holes may provide
interesting material for future work.

\bibliographystyle{unsrt}
\bibliography{tom}

\end{document}